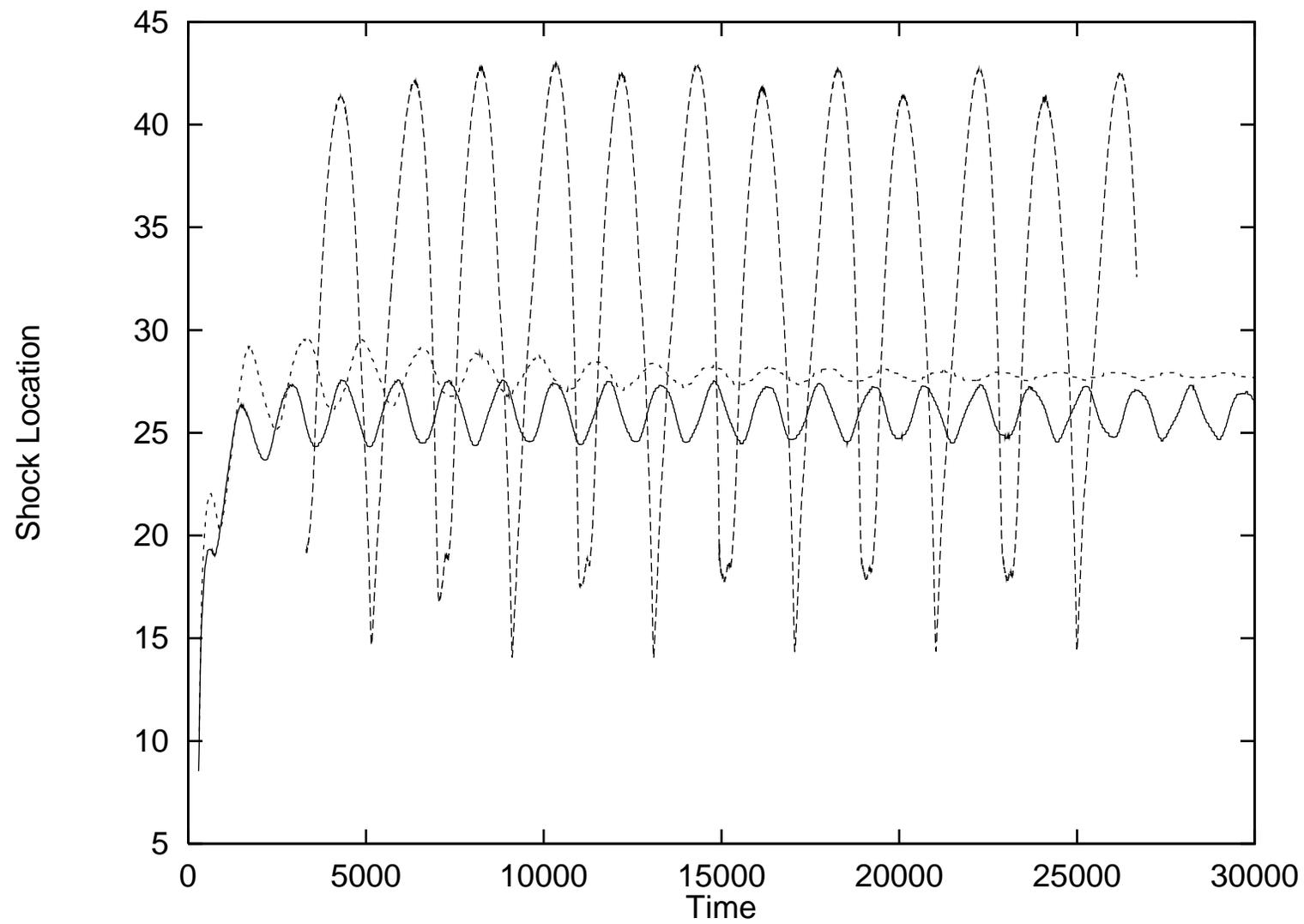

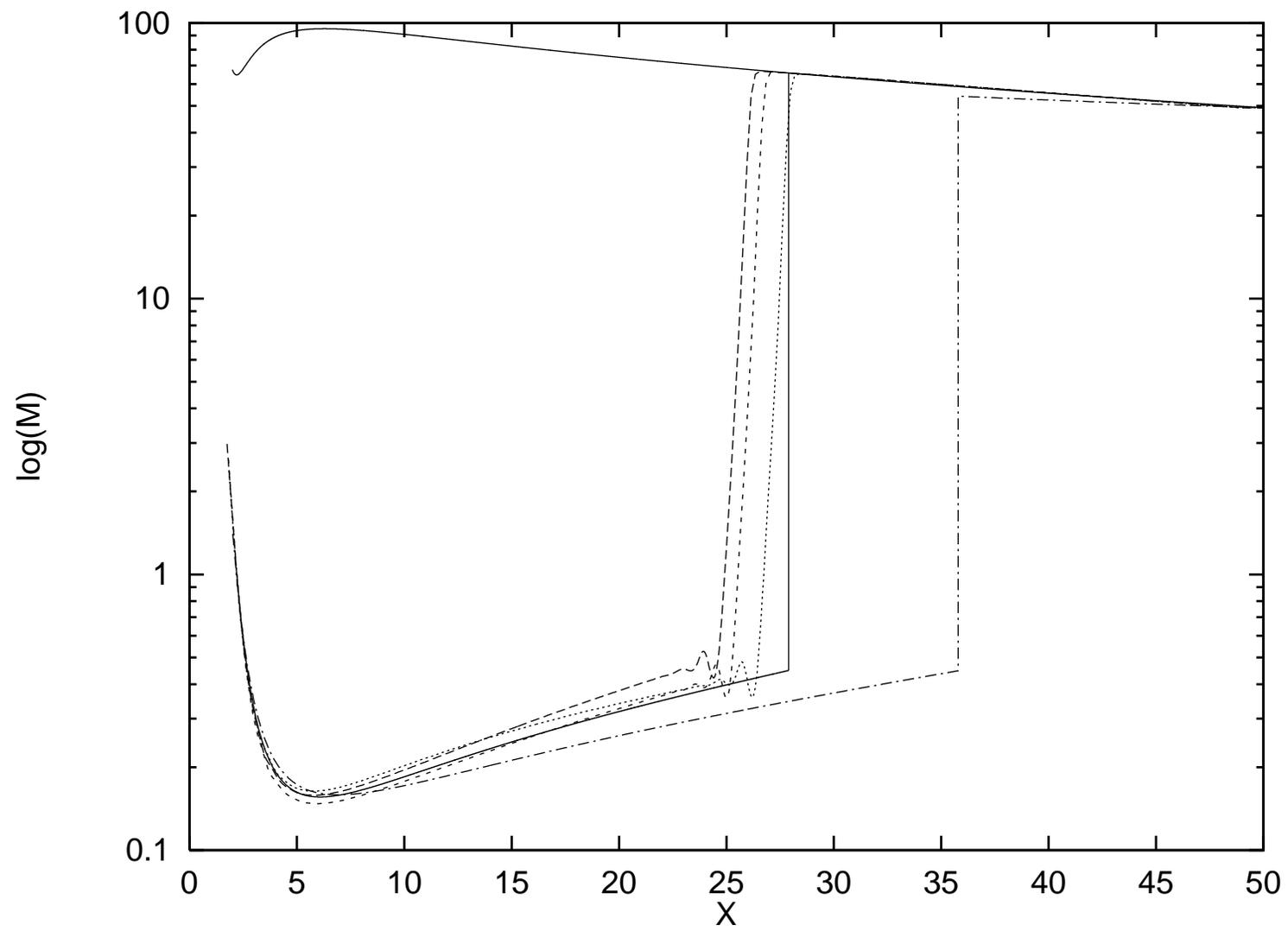

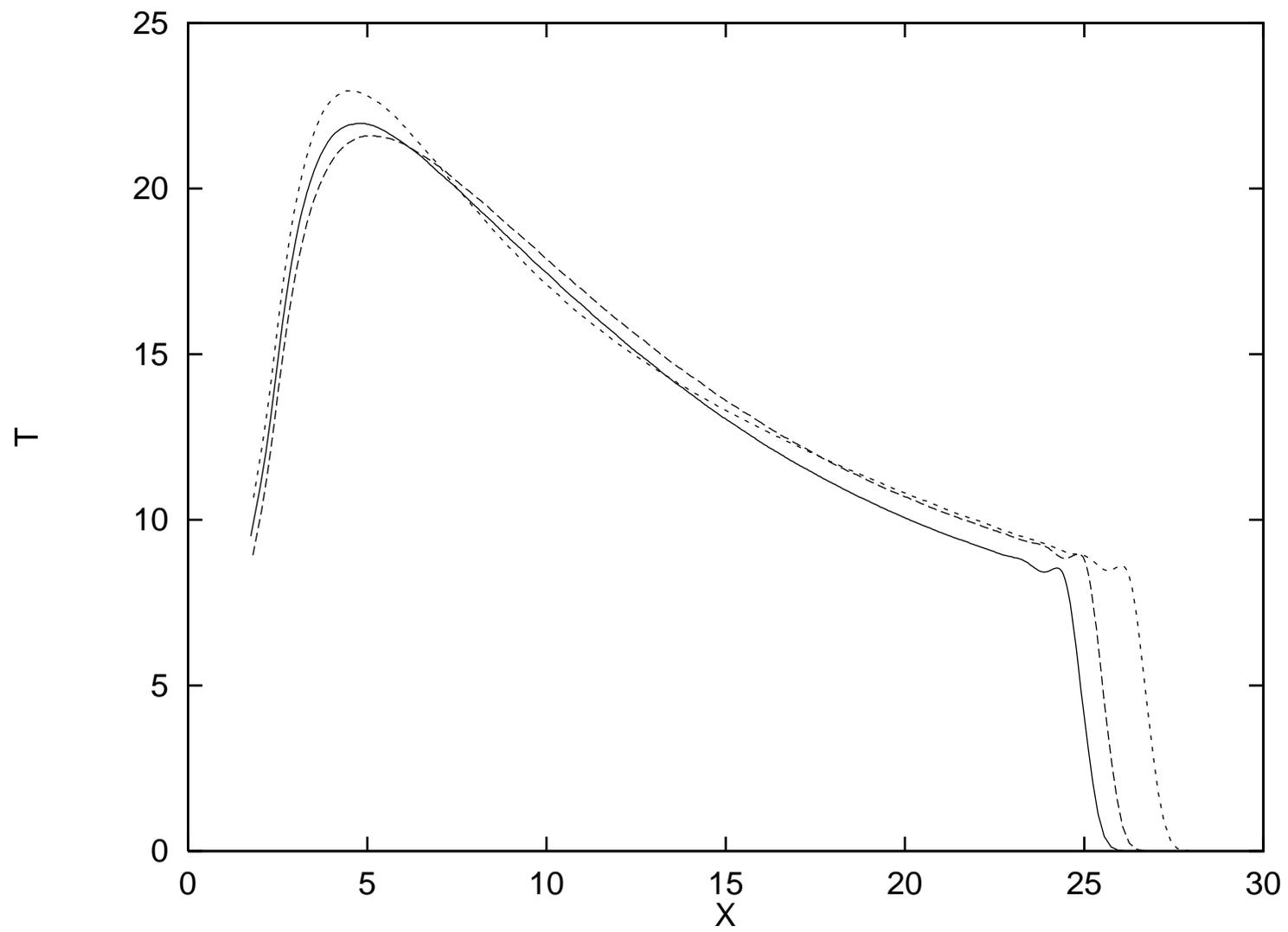

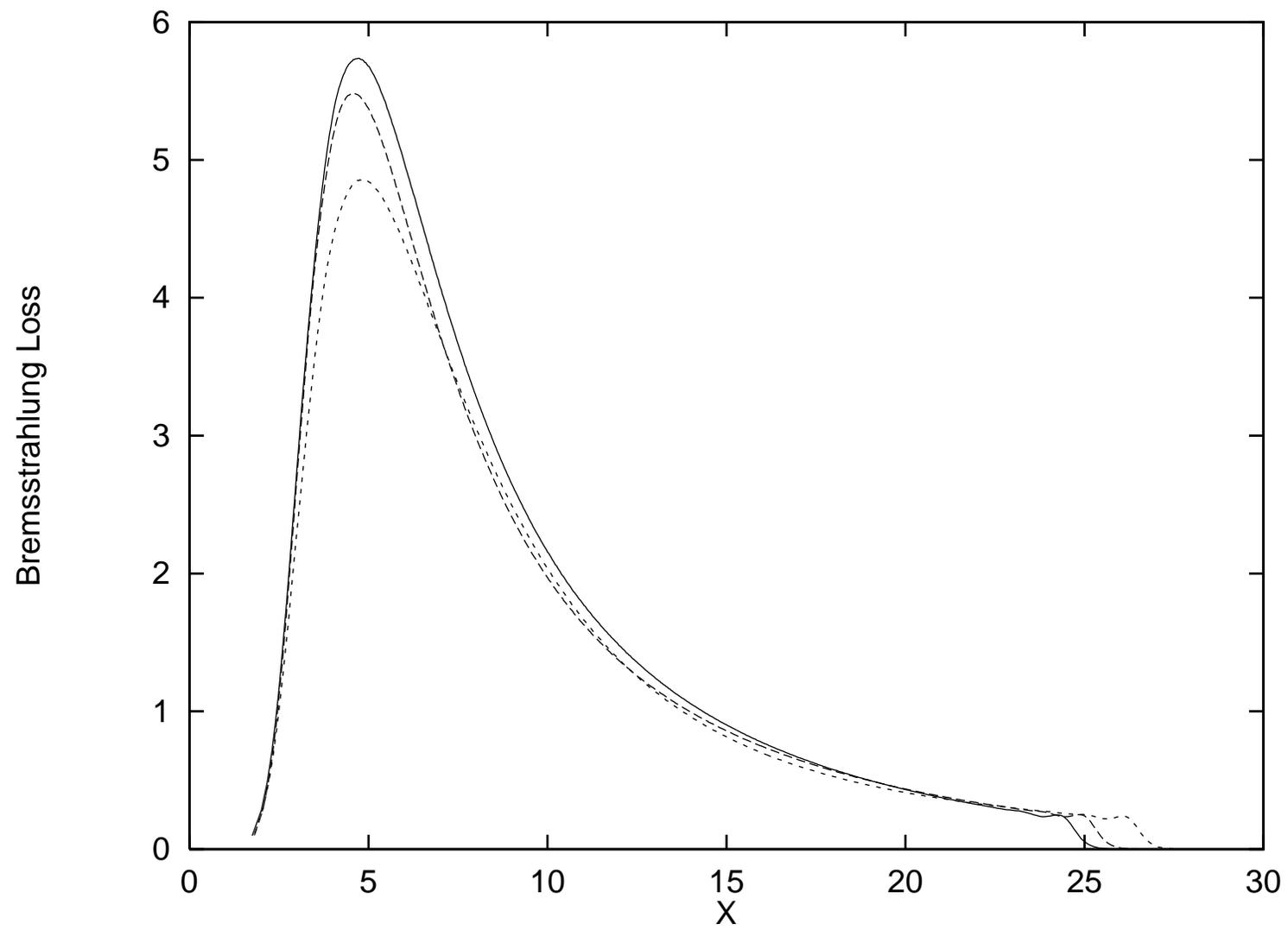

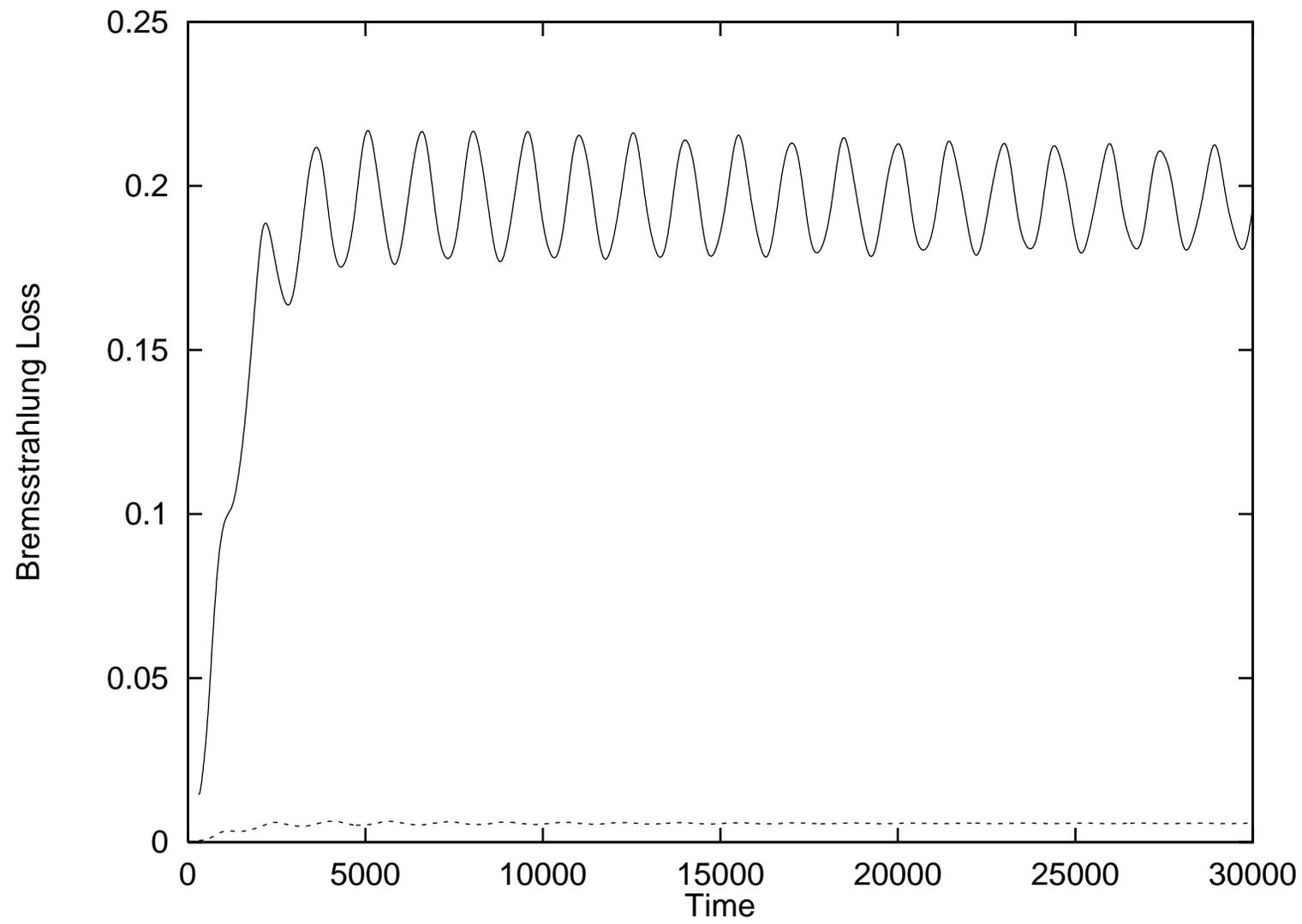

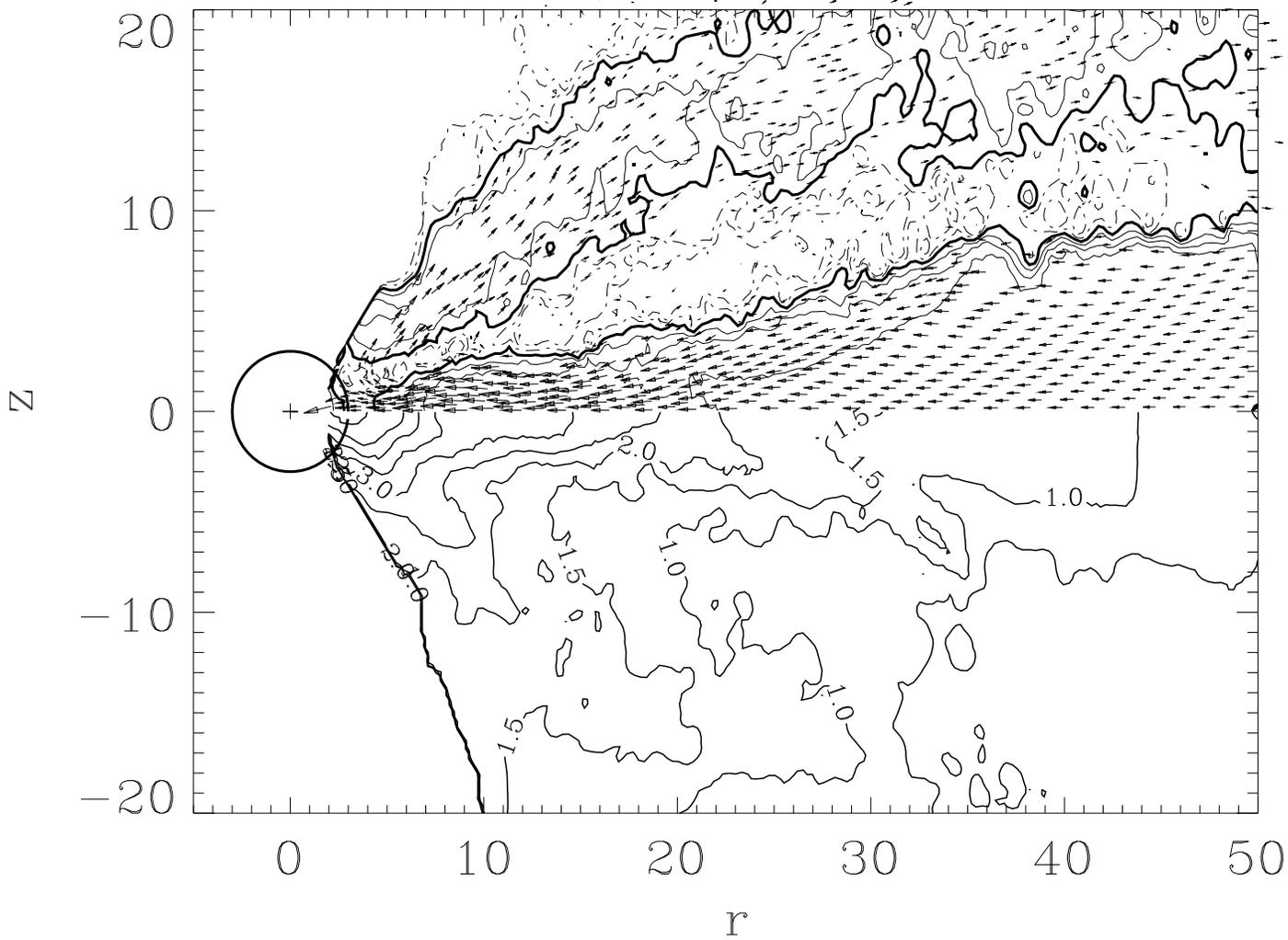

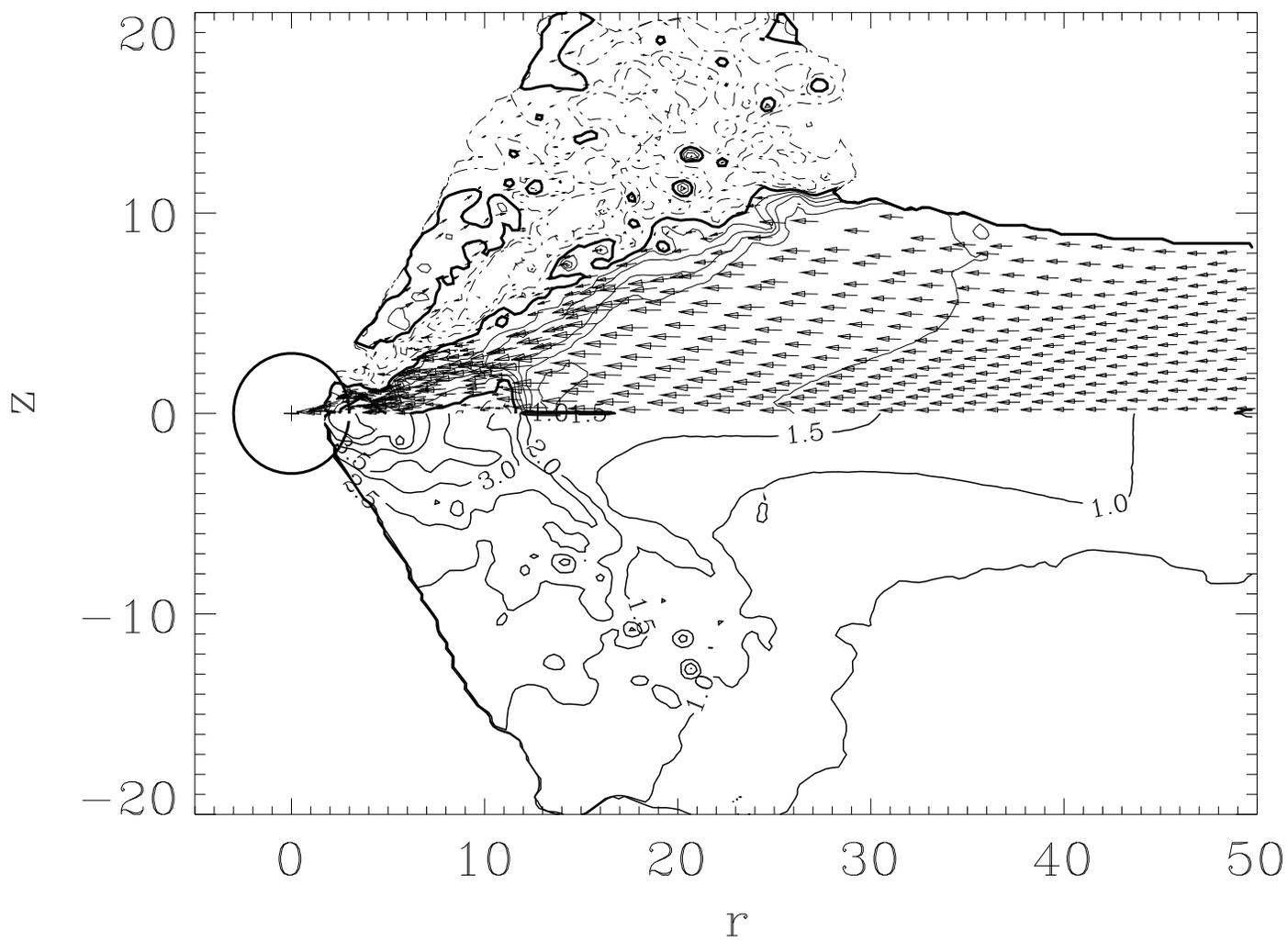

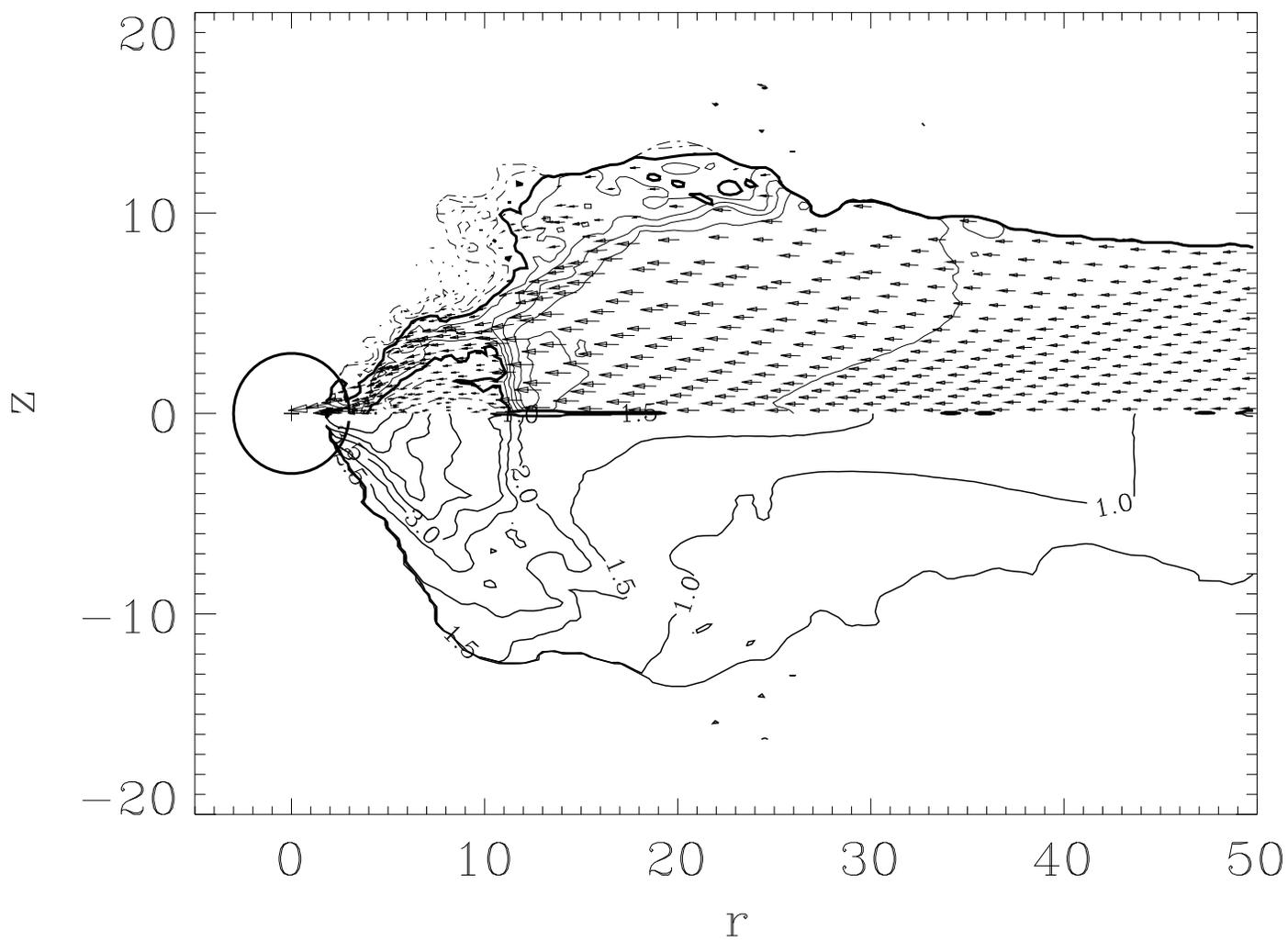

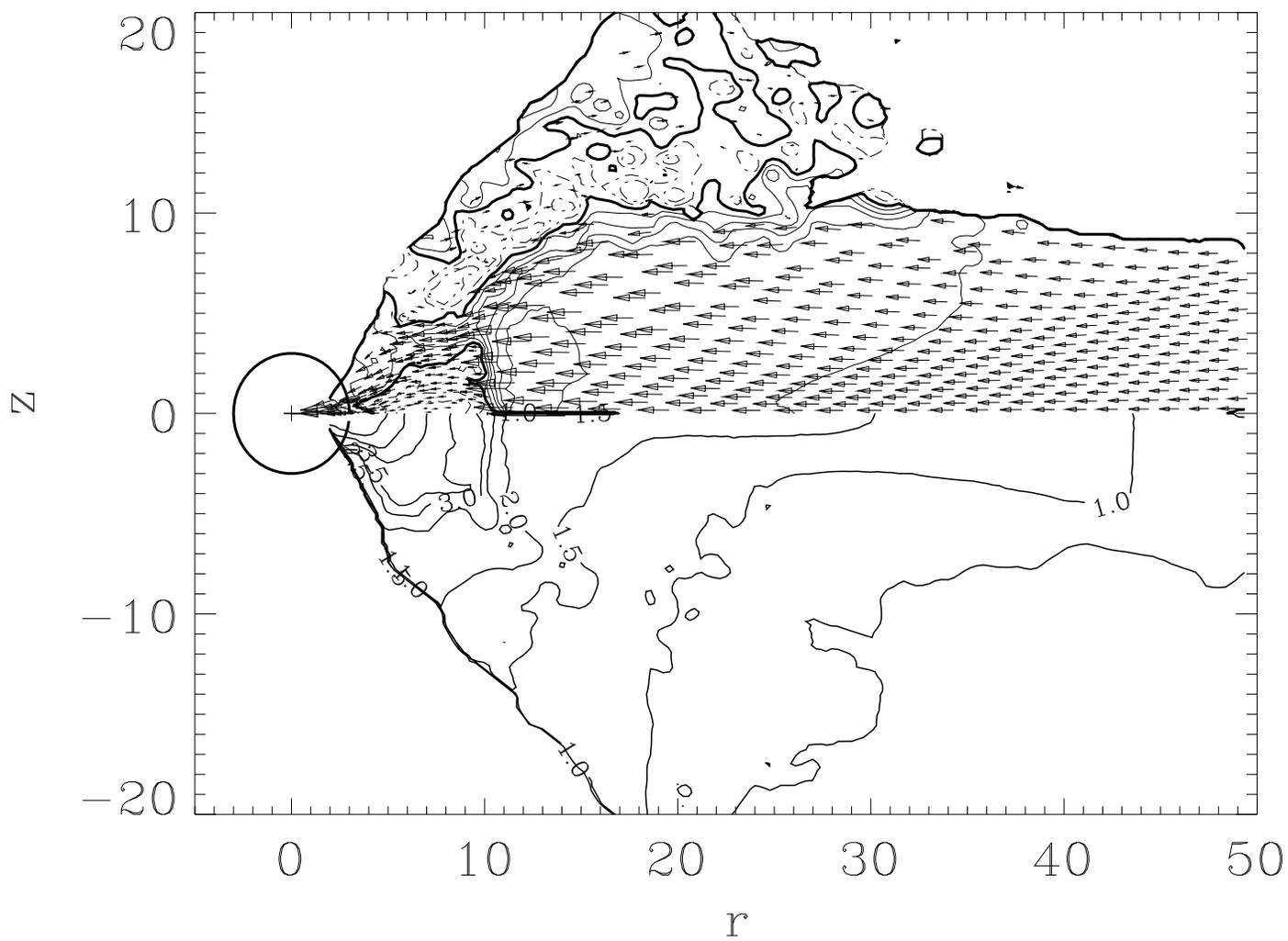

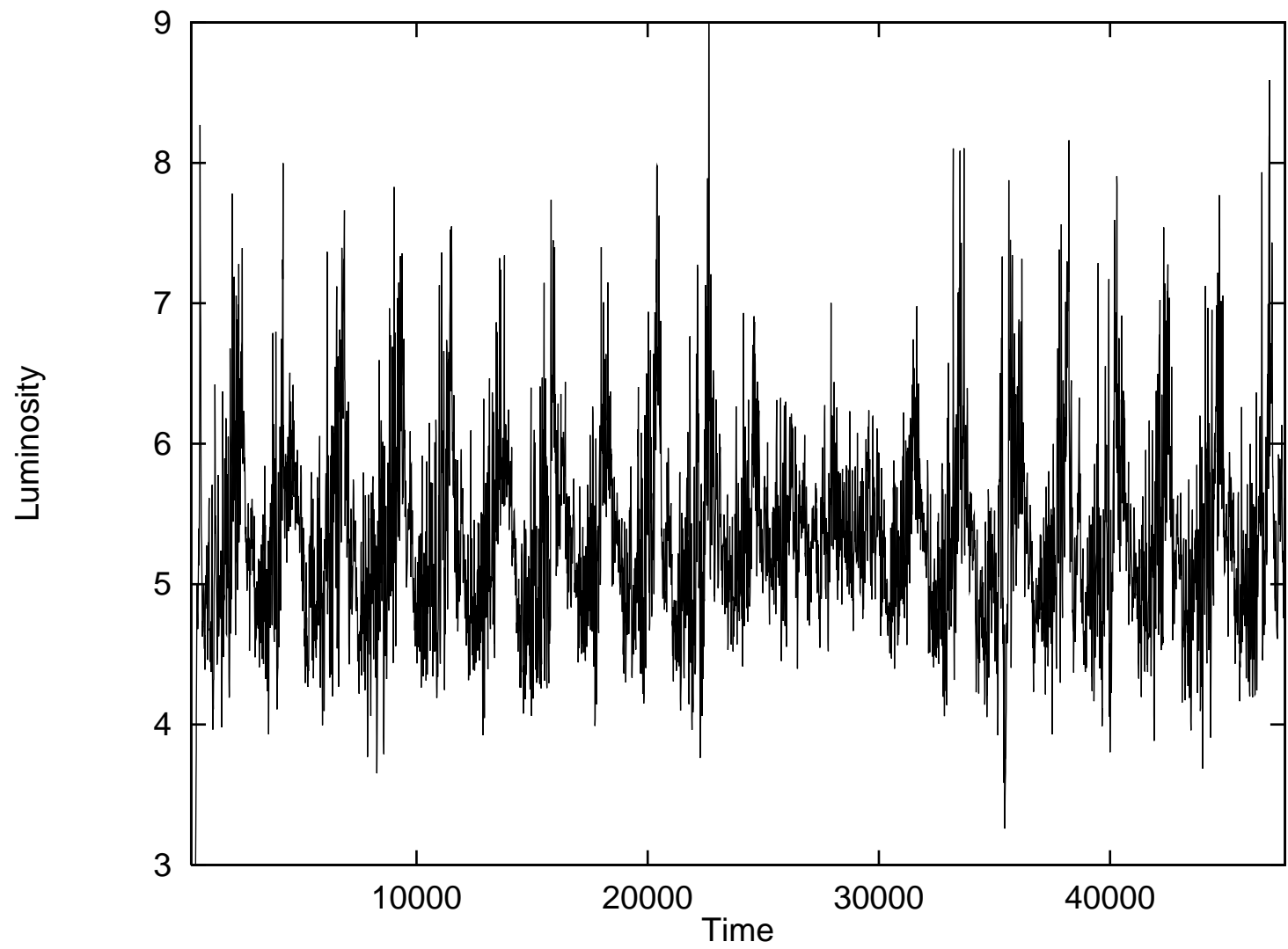

# Resonance Oscillation Of Radiative Shock Waves In Accretion Disks Around Compact Objects

Diego Molteni
University of Palermo, 90100 Palermo, Italy

Hanno Sponholz*
Institut für Theoretische Astrophysik & Interdisziplinäres Zentrum
für Wissenschaftliches Rechnen der Universität Heidelberg
Im Neuenheimer Feld 561
D – 69120 Heidelberg

Sandip K. Chakrabarti
Code 665, Goddard Space Flight Center, Greenbelt, MD, 20771[1]
and
Tata Institute Of Fundamental Research, Bombay, 400005[2]

August 4, 1995

## Abstract

We extend our previous numerical simulation of accretion disks with shock waves when cooling effects are also included. We consider bremsstrahlung and other power law processes: $\Lambda \propto T^\alpha \rho^2$ to mimic cooling in our simulation. We employ *Smoothed Particle Hydrodynamics* technique as in the past. We observe that for a given angular momentum of the flow, the shock wave undergoes a steady, radial oscillation with the period is roughly equal to the cooling time. Oscillations seem to take place when the disk and cooling parameters (i.e., accretion rate, cooling process) are such that the infall time from shock is of the same order as the post-shock cooling time. The amplitude of oscillation could be up to ten percent of the distance of the shock wave from the black hole when the black hole is accreting. When the accretion is impossible due to the centrifugal barrier, the amplitude variation could be much larger. Due to the oscillation, the energy output from the disk is also seen to vary quasi-periodically. We believe that these oscillations might be responsible for the quasi periodic oscillation (QPO) behaviors seen in several black hole candidates, in neutron star systems as well as dwarf novae outbursts such as SS Cygni and VW Hyi.

*Subject Headings*: black holes — stars: accretion —- stars: winds — shocks waves – radiation hydrodynamics - numerical simulations

[1] NRC Senior Research Associate
[2] Permanent Address
[*] Present Address: Max-Planck-Gesellschaft, Arbeitsgruppe Gravitationstheorie a. d. Friedrich-Schiller-Universität Jena, Max-Wien-Platz No. 1, D – 07743 Jena
e-mails: molteni@gifco.fisica.unipa.it; chakraba@tifrvax.tifr.res.in

# 1. INTRODUCTION

It is becoming increasingly clear that if in active galaxies accretion takes place in the form of accretion disks, then these disks may not be similar to the standard 'Shakura-Sunyaev' type thin accretion disks models found most suitable for binary systems. The presence of virtually simultaneous variabilities in optical and UV in several objects such as NGC 5548 (Clavel et al. 1990, 1992 and Peterson et al. 1991), NGC 4151 (Perola et al. 1986) and temporal variation of the line emissions in broad line radio galaxies, such as ARP 102B, 3C390.3 (Miller & Peterson, 1990; Veilleux & Zheng, 1991; Zheng, Veilleux & Grandi, 1991; Chakrabarti & Wiita, 1994) are examples of the fact that the some disks are of completely different type and a significant contribution to the emitted spectra could be made by the shock waves in the disks. These shock waves could either be pressure and angular momentum supported (Chakrabarti, 1989 hereafter C89; Chakrabarti, 1990 hereafter C90) in the case of inflows which include low but significant angular momentum, or only pressure supported (Chang & Ostriker, 1985; Kazanas & Ellison, 1986; Babul, Ostriker & Meszaros, 1993) when the flow is spherical. Whether or not accretion disks may have a shock or simply accrete onto the black hole through inner sonic point could be determined from the viscosity, accretion rate and angular momentum of the flow. This unified classification of the global solution is discussed in C89.

In our previous work (Chakrabarti & Molteni 1993; hereafter CM93), we presented hydrodynamic simulation of the standing shock waves in thin accretion disks. We showed in particular, that standing shock waves are formed in inviscid accretion disks of low angular momentum $\lambda \leq \lambda_{mb}$, where $\lambda_{mb}$ is the marginally bound angular momentum. Subsequently, Molteni, Lanzafame & Chakrabarti (1994; hereafter MLC94), studied numerical simulation of shock waves in quasi-spherical accretion disks around a Schwarzschild black hole and verified that the results agree very well with the vertically averaged solution (C89). Sponholz and Molteni (1994) studied shock formation around a Kerr black hole and find differing shock locations in co- and contra-rotating flows. When angular momentum supply at the outer edge of the disk is large, a significant viscosity must be present in the disk in order that accretion may take place. In an important simulation, Chakrabarti & Molteni (1995, hereafter CM95) pointed out that when viscosity is low enough (in the language of $\alpha_{vis}$ parameter of Shakura-Sunyaev 1973; $\alpha_{vis} < \alpha_{vc}$), a weaker shock may still form farther away from the black hole. For a higher viscosity ($\alpha_{vis} > \alpha_{vc}$), the disk is shock-free and is similar to a standard Keplerian disk at larger distance while passing through an inner sonic point. These results are consistent with the observation of the most general solution of the viscous transonic disks (C90) where the critical viscosity parameter ($\alpha_{vc}$) for isothermal viscous disk is computed. Recent works of 'newly discovered advection dominated' flows (Narayan & Yi, 1994, Narayan & Yi, 1995), which are the generalization of shock-free global solutions of C90, (for cooling processes not satisfying isothermality condition) miss the shock solutions completely because of the restrictive assumption of self-similarity and the choice of the Newtonian potential. Thus the observations derived from our more general consideration will not be applicable to such solutions.

In the present paper, we deviate from our previous studies of simulation in two ways: (a) we use an adiabatic index $\gamma = 5/3$ (instead of $\gamma = 4/3$) to mimic a gas with a low radiation pressure and (b) we include a a power-law cooling process $\Lambda \propto \rho^2 T^\alpha$ in the disk. We extensively study the case of $\alpha = 0.5$ (i.e., bremsstrahlung) and discover a completely new physical effect not discussed so far in the literature in the context of transonic accretion disks around a black hole. We show that the shock wave undergoes a steady quasi-periodic oscillation in radial direction with an amplitude roughly about five to ten percent of the steady state shock distance from the black hole and the time period similar to the cooling time. The oscillation causes the luminosity of the disk to vary quasi-periodically as well. We find a resonance behavior in the oscillation characteristics in the accreting disk: if the disk parameters (such as the accretion rate and angular momentum) and the cooling parameters (such as $\alpha$ and efficiency of cooling relating to heating) are such that the infall time from the shock location is comparable to the radiative cooling time, then the shock oscillates, otherwise only the steady state shock remains.

One of the motivations of our study stems from the results of the Langer, Chanmugam & Shaviv (1981, 1982) who showed that shocks in accretion columns of white dwarfs and neutron stars exhibit oscillational instability in presence of a power law cooling. Subsequent work of linear stability analysis (Chevalier & Imamura 1982) shows that such an oscillation is expected. These results were further generalized by several workers by including effects of cyclotron cooling (Chanmugam, Langer, & Shaviv, 1985) in accretion onto magnetized white dwarfs who found that the cooling at a field strength $B \sim 30 \ mG$ is sufficient to stabilize the oscillation. Wolff, Gardner & Wood (1989) studies one and two temperature models including bremsstrahlung, Compton cooling and cyclotron cooling and Wu, Chanmugam & Shaviv (1992) studies oscillations in continuously excited accretion flows with a boundary condition relevant for white dwarfs. Other works discussed analogies between the instabilities in shocks due to cooling and exothermic reactions (e.g. Falle, 1995 and references therein). Our results are new in the following respects: Contrary to white dwarf or neutron star, a black hole does not have a 'hard' surface. However, in presence of weaker dissipation, the centrifugal barrier experienced by the flow can be strong enough to act as a 'hard' surface where eventually a standing shock wave is formed. Thus, the question arises: would these shock waves oscillate in the manner it does on white dwarf surface? We find that they do and we believe that they may be responsible for the QPO behaviors observed in many black hole candidates. Secondly, we found that the oscillation is actually of 'resonance' type, namely, it occurs only the the cooling time scales roughly match with the infall time scale. If the cooling time is very short the flow rapidly adjusts to the cooler disk, but if the cooling time is very large, the thermal energy is advected along with the flow. As in earlier three-dimensional simulation (Molteni, Lanzafame & Chakrabarti, 1994), we find that the post-shock region forms a corona which could be the site of hard X-ray and $\gamma$-ray production.

The general theory of the study of the shock waves in accretion disks are presented in C89, C90 and CM93 and we shall not repeat them here. In the next Section, we shall present the equations used in our numerical simulations. In §3, we present the way bremsstrahlung is implemented in our axisymmetric SPH code. In §4, we present numerical solutions and compare with the analytical results. Finally, in §5, we make concluding remarks.

## 2. MODEL EQUATIONS

We assume a rotating, axisymmetric, accretion flow around a black hole. We take Newtonian model for the non-rotating central compact object as given in terms of the Paczyński & Wiita (1980) potential which is found to be accurate enough for astrophysical purposes. The internal energy $e$ of the disk is defined as $P = \rho e(\gamma - 1)$. $P$ and $\rho$ are the isotropic pressure and the matter density respectively, $\gamma = 5/3$ is the adiabatic index. Only single temperature model is considered here so that electrons and protons have the same temperature (two temperature disk solutions will be presented elsewhere). We presently assume that the flow is inviscid, i.e., the specific angular momentum $\lambda$ is constant everywhere. We use the density of the disk at the outer edge $\rho_{ref} = \rho_0$, the velocity of light $v_{ref} = c$ and the Schwarzschild radius $x_{ref} = R_g = 2GM/c^2$ of the black hole mass $M$, as the reference density, velocity and distance respectively. Thus, for example, the unit of time is $t_{ref} = x_{ref}/v_{ref} = R_g/c$, the unit of angular momentum is $\lambda_{ref} = x_{ref} v_{ref} = cR_g$, the unit of mass is $M_{ref} = \rho_{ref} x_{ref}^3 = \rho_{ref} R_g^3$, and the unit of mass accretion rate is $\dot{M}_{ref} = \rho_{ref} x_{ref}^3 / t_{ref} = \rho_{ref} R_g^2 c$.

First, we consider one-dimensional equations. Our choice is restricted by the following constraints. Whereas it is easy to obtain analytical solutions for disks of constant thickness (e.g., CM93) or disks in vertical equilibrium (e.g., C89), it is very difficult to obtain solutions for full three-dimensional flows. An axisymmetric numerical code can easily test analytical solution of one-dimensional axisymmetric flow of constant thickness and it can also perform a fully three-dimensional simulation without taking resort to the vertical equilibrium assumption. Thus, we solve equations of constant thickness (rather than vertically averaged model) first to verify that the time dependent code is satisfactory. Subsequently, we solve fully three-dimensional equations as presented in §3. The time-dependent equations which we solve are:

(a) The energy equation:
$$\frac{\partial e}{\partial t} = -v \frac{\partial e}{\partial x} - \frac{P}{\rho} \frac{\partial v}{\partial x} - \zeta_{1/2} \rho e^\alpha \quad . \tag{1a}$$

(b) The radial momentum equation:
$$\frac{\partial v}{\partial t} + v \frac{\partial v}{\partial x} + \frac{1}{\rho} \frac{\partial P}{\partial \rho} + \psi'(x) = 0 \quad . \tag{1b}$$

(c) The continuity equation:
$$\frac{\partial \rho}{\partial t} + \frac{1}{x} \frac{\partial}{\partial x} (\rho v x) = 0 \quad . \tag{1c}$$

Here, $x$ is the radial coordinate, $\psi(x)$ is the effective potential energy $\psi(x) = g(x) + \lambda^2/2x^2$ with $g(x)$ as the radial force potential, which, in the pseudo-Newtonian model, takes the form: $g(x) = -1/2(x-1)^{-1}$, $e$ is the internal energy density $e = \frac{P}{(\gamma-1)\rho}$, $v$ is the radial component of velocity and $a$ is the adiabatic sound speed: $a^2 = \frac{\gamma P}{\rho}$ Primes denote a derivative with respect to $x$. $\zeta_\alpha$ is the non-dimensional bremsstrahlung loss coefficient,

$$\zeta_{1/2} = \frac{j\rho_{ref}x_{ref}T_{ref}^{1/2}}{c^3 m_p^2} \qquad (2)$$

and

$$T_{ref} = \frac{c^2 m_p \mu(\gamma - 1)}{k} \qquad (3)$$

where, $\mu = 0.5$ and $j = 1.4 \times 10^{-27}$ c.g.s. unit for ionized hydrogen (Allen 1973), $m_p$ is the mass of the proton, and $k$ is the Boltzmann constant. The subscript $1/2$ of $\zeta$ reflects that we are using the cooling law $\Lambda = \zeta_{1/2}\rho^2 T^\alpha$ with the constant $\zeta_{1/2}$ exactly same as in bremsstrahlung case ($\alpha = 0.5$).

In the steady state, the above equations could be solved quite easily in one dimension by assuming $\partial/\partial t = 0$ and re-arranging them first in the form:

$$\frac{dv}{dx} = \frac{\frac{d\psi}{dx} - \frac{a^2}{x} + \frac{2\zeta_{1/2}a^{2\alpha}\rho}{3v}}{a^2/v - v} \qquad (4a)$$

which along with the following equations,

$$v\frac{dv}{dx} + \frac{2a}{\gamma}\frac{da}{dx} + \frac{a^2}{\gamma\rho}\frac{d\rho}{dx} + \frac{d\psi}{dx} = 0 \qquad (4b)$$

and

$$\frac{1}{v}\frac{dv}{dx} + \frac{1}{\rho}\frac{d\rho}{dx} + \frac{1}{x} = 0. \qquad (4c)$$

Detailed solution properties of the sonic points and shocks in presence of cooling effects are beyond the scope of this paper and will be discussed elsewhere.

A complete solution which includes a shock wave 'glues' together two integral curves passing through the outer and inner sonic points respectively at the shock location $x_s$ determined by the Rankine-Hugoniot relation at the shock (Chakrabarti, 1990). Assuming the shock to be infinitesimally thin and non-dissipative at the shock itself, these relations are of (a) the conservation of the energy flux, (b) the conservation of the momentum flux and (c) the conservation of the mass flux, respectively across the shock wave. When these relations are combined together, one obtains the so-called 'Mach number relation' as:

$$\frac{M_-^2 + 3}{(\frac{1}{M_-} + \gamma M_-)^2} = \frac{M_+^2 + 3}{(\frac{1}{M_+} + \gamma M_+)^2} = \text{Constant} \qquad (5)$$

where, $-$ and $+$ signs are used to denote quantities in the pre-shock and post-shock regions. This relation is used to obtain the location of the shock wave in a steady state. Before we present the analytical and numerical solutions, we present a few sentences about the way the cooling effect is implemented in smoothed particle hydrodynamics simulations (SPH).

## 3. SPH: IMPLEMENTATION OF THE COOLING LAW

To find non-steady-state solutions of the classical one-dimensional hydrodynamical equations 1(a-c) of a compressible inviscid fluid we apply the Smoothed Particle Hydrodynamics (SPH) method, specially adopted for axial symmetric configurations as in our earlier works. The procedure presented here is valid for thick $(r, z)$ accretion as well and is described in detail in Molteni & Sponholz (1994) as well as in CM93. While applying to the thin disks (1a-c), one has to assume a constant vertical thickness and inject matter only on the equatorial plane.

The basic point for the cylindric geometry approach is to assume the interpolating Kernel $W$ which is a function of cylindrical radial coordinate $\vec{r} = \vec{r}(x, z)$ and $k$-th particle of mass $m_k$ as,

$$m_k = 2\pi \rho_k x_k \Delta \vec{r}_k \qquad (6)$$

Any smooth function $A(\vec{r}_i)$ at $\vec{r}_i$ is defined as,

$$A(\vec{r}_i) = \int A(\vec{r}) W(\vec{r} - \vec{r}_i; h) \frac{2\pi \rho x}{2\pi \rho x} d\vec{r} \approx \sum_k \frac{m_k}{2\pi x_k \rho_k} A(\vec{r}_k) W(\vec{r}_k - \vec{r}_i; h), \qquad (7)$$

$h$ being the particle size. Thus, for example, for the density at each point, we have the simple expression that identically satisfy the continuity equation in the cylindrical coordinate:

$$\rho(\vec{r}_i) \approx \sum_k \frac{m_k}{x_k} W(\vec{r}_k - \vec{r}_i; h) \tag{8}$$

The equations of motion to be solved consist of the radial momentum equation:

$$\left(\frac{Dv_x}{Dt}\right)_i = -\sum_k \frac{m_k}{x_k}\left(\frac{P_i}{\rho_i^2} + \frac{P_k}{\rho_k^2} + \Pi_{ij}\right)\frac{\partial W_{ik}}{\partial x_i} + \frac{\lambda^2}{x_i^3} - \frac{1}{2(r_i-1)^2}\frac{x_i}{r_i}, \tag{9a}$$

the vertical component of the momentum equation.

$$\left(\frac{Dv_z}{Dt}\right)_i = -\sum_k \frac{m_k}{x_k}\left(\frac{P_i}{\rho_i^2} + \frac{P_k}{\rho_k^2} + \Pi_{ij}\right)\frac{\partial W_{ik}}{\partial z_i} - \frac{1}{2(r_i-1)^2}\frac{z_i}{r_i}. \tag{9b}$$

The energy equation describes the behavior of the thermal energy per unit mass and contains the cooling-term $\Lambda_i = \zeta_{1/2}\rho_i^2(\mathcal{E}_i)^\alpha$:

$$\left(\frac{D\mathcal{E}}{Dt}\right)_i = -\frac{1}{2}\sum_k \frac{m_k}{x_k}\left(\frac{P_i}{\rho_i^2} + \frac{P_k}{\rho_k^2} + \Pi_{ij}\right) - \frac{\Lambda_i}{\rho_i} \tag{9c}$$

To mimic numerically the kinematic dissipation, the artificial viscosities:

$$\Pi_{ij} = \frac{\alpha\mu_{ij}\bar{c}_{ij} + \beta\mu_{ij}^2}{\bar{\rho}_{ij}}$$

$$\tilde{\mu}_{ij} = \frac{x_i v_{xi} - x_j v_{xj}}{x_i(l_{ij}^2 + \eta^2)} + \frac{(v_{zi} - v_{zi})(z_i - z_j)}{l_{ij}^2 + \eta^2},$$

$$l_{ij}^2 = (x_i - x_j)^2 + (z_i - z_j)^2, \quad \eta = 0.1h^2$$

with the abbreviations for density $\bar{\rho}_{ij}$ and sound speed $\bar{a}_{ij}$ are used (Monaghan 1992):

$$\bar{\rho}_{ij} = \frac{\rho_i + \rho_j}{2} \quad \text{and} \quad \bar{a}_{ij} = \frac{a_i + a_j}{2},$$

$$\eta = 0.1h^2,$$

$\alpha$ and $\beta$ control the amount of the artificial viscosity, necessary to reduce oscillations in shock transitions.

Thus, the implementation of the SPH in cylindrical code is similar in every respect to the original SPH code in Cartesian coordinate, except that the mass of a particle appears divided by its axial distance and the relative velocity $v_k - v_i$ between $k$-th and $i$-th particles must be replaced by a $(x_k v_k - x_i v_i)/x_i$ term (Chakrabarti & Molteni, 1993; Molteni & Sponholz, 1994).

## 4. RESULTS AND PHYSICAL INTERPRETATIONS

Below, we first present analytical and numerical solutions in thin accretion disks with constant thickness. Subsequently, we present more realistic solutions for three-dimensional accretion disks. In the next sub-Section, we provide physical interpretation of these solutions.

### 4.1 Results in Thin Accretion Disks

While computing the results we have fixed our attention keeping an application to the active galaxies in mind. We have chosen the mass of the black hole to be $M = 10^8 M_\odot$. We choose the specific angular momentum (in units of $2GM/c$) to be $\lambda = 1.9$ which lies between marginally stable and marginally bound values. We inject matter at the outer edge of the disk $x = 50 R_g$ with $v \sim 0.14$ and $a \sim 0.0024$. Matter reaching the inner boundary (chosen here to be at $x = 1.5$) are removed.

To obtain an analytical solution, we choose a typical set of parameters (such as, the density $\rho$, the velocity $v$, the sound speed $a$ and the specific angular momentum $\lambda$) at the boundary and solve for $\rho(x)$, $v(x)$ and $a(x)$ using equations 4(a-c). We have obtained steady state solutions using fourth order Runge-Kutta equations.

Figure 1 compares the shock locations as a function of time for different value of $\alpha$. In all the cases, the density at the outer edge as $1.85 \times 10^{-15}$ gm cm$^{-3}$. At this density, with a uniform vertical thickness of one Schwarzschild radius, the mass accretion rate is given by $\dot{M} = 0.14 M_\odot$ yr$^{-1}$ which is about seventy percent of of the Eddington rate for a black hole of mass $10^8 M_\odot$. The solid curve is for $\alpha = 0.5$, and the long dashed and short dashed curves are for $\alpha = 0.4$ (largest amplitude) and $\alpha = 0.75$ (smallest amplitude) respectively. A large number of important conclusions could be drawn: (a) for bremsstrahlung cooling ($\alpha = 0.5$), the shock oscillates in a very stable manner. The shock location $x_s$ varies by about five percent of the mean location. The period of oscillation is $T_{1/2} \sim 1500$ unit, which is equal to $\sim 1.5 \times 10^6$s or 17.4d. (b) For $\alpha = 0.75$, the oscillation of the shock damps out completely, and the mean location is higher than the average $x_s$ for $\alpha = 0.5$. In both (a) and (b), matter is found to accrete through the inner sonic point. (c) For $\alpha = 0.4$, the amplitude is very high. In this case, no accretion takes place. The flow in the post-shock region cools down so rapidly that the pressure is not sufficient to push matter across the centrifugal barrier. Presence of viscosity is expected to reduce this barrier as well as the amplitude of oscillation. The time period of oscillation is about $T_p \sim 1978$ unit, i.e., 22.9d for a $10^8 M_\odot$ black hole. In all these simulations the numerical length-scale (interpolation-length $h$) size is chosen to be 0.25 which is sufficiently small compared to the amplitude of oscillation. Therefore we believe that these oscillations are physical. The number of particles with the region of integration was about 200 in our simulation.

In order to show that these oscillations are not artifacts of our simulation, we have tested the result in numerous ways, particularly by varying the particle size $h$ and viscosity parameters. We find no difference of any significance. More importantly, we tested these solutions against analytical steady state solution using same method as in C89 and C90. In Fig. 2 we present one such test result where both the time dependent (Eqs. 1a-c) and time independent (Eqs. 4a-c) solutions are shown. We have chosen here $\alpha = 0.5$. The solid curves (supersonic and subsonic branches) represent the steady state result, while the long-dashed, short-dashed and dotted curves are the time dependent solutions at different phases of oscillation. The time dependent solutions, particularly in the subsonic branch, are seen to oscillate around the steady state solution. For comparison, we present the dash-dotted curve which is the steady state solution with no cooling effects present. In the pre-shock region, temperature and densities are low and cooling effects are negligible. In the post-shock region, cooling effects are very strong. In Fig. 3(a-b) we show how the (a) temperature (in units of $10^9$K) and (b) total bremsstrahlung loss (in arbitrary units) vary with distance from the black hole, as well as at various phases of oscillation. It is important to note that the whereas the temperature varies significantly throughout the post-shock region, maximum variation of the loss takes place much closer to the black hole at $X \sim 5$, because of higher densities (Of course, significant variation occurs at the mean location of the shock, due to the movement of the shock). When the shock is at its outer most position (short dashed curve) the bremsstrahlung loss in minimum. However, the net loss is still a strongly varying function of time. Figure 4 shows the variation of net loss with time. The solid curve is for $\alpha = 0.5$ and the dashed curve is for $\alpha = 0.75$. The variation of bremsstrahlung luminosity is about 15 percent. We do not find any steady accreting solution for the above accretion rate (i.e. $\dot{M} = 0.6\dot{M}_{Edd}$) when $\alpha = 0.4$. The time dependent solution shows larger variations in the amplitude of oscillation (Fig. 1). We shall comment on the implications of this results in next Sections.

## 4.3 Simulations of Thick Accretion Disks

So far, we have presented the behavior of shock waves in a thin accretion disk. Presently, we show behaviors of axisymmetric three dimension axisymmetric flows whose SPH implementation is given in Eqs. 9(a-c). In these simulations we have assumed the disk to be symmetric with respect to the equatorial plane. Solutions with-out the latter boundary conditions show very interesting behavior and we shall discuss about them elsewhere (Sponholz, Molteni & Chakrabarti 1995, in preparation). In Fig. 5(a-d), we show results of three dimensional simulations. The parameters at the outer edge ($X = 50$) are chosen to be: density $\rho = 4 \times 10^{-14}$ gm c$^{-1}$, infall velocity $u = 0.126$ and the sound velocity $a = 0.04$. This corresponds to an accretion rate of $\dot{M} = 10.8 M_\odot$ yr$^{-1}$, or, about $\dot{M} \sim 47 \dot{M}_{Edd}$. The specific angular momentum chosen is $\lambda = 1.75$. With this low angular momentum, matter is still optically thin except extremely close to the black hole which is much inside the shock region and are not expected to influence our solution. The number of particles in the simulation is about 2000. In the upper half of each panel we show velocity vectors (whose sizes are proportional to the actual velocities), the thick curves are the the contours of constant Mach number $M = 1$. The thin lines presents on both (sub- and supersonic) sides represents contours of constant Mach numbers $\log(M) = -1.0, -0.5,$ and $0.5, 1.0, 1.5, ...3.$, whereas the dashed contours mark the subsonic matter. In the lower half of each panel we show contours of constant density in units of the density at the outer edge. The circle around the origin is of three Schwarzschild radius, the size of the marginally stable orbit. In Fig. 5a, we produce the solutions without bremsstrahlung. A very weak shock is formed very close to the black hole $X \sim 5$ and a strong wind is also produced from the high entropy, post-shock flows very similar to what is observed in MLC94. The solution roughly remains the same with time.

When bremsstrahlung loss is included, the situation is changed dramatically. The shock is formed at around $X = 10$ and a strong corona with outflowing wind is also produced. As the shock oscillates, the size of the corona also changes. The net luminosity due to bremsstrahlung effect is also varied by 15 to 20 percent. The time period of oscillation is about 2200. Figure 6, shows the oscillation of the emitted radiation from the three dimensional simulations. The oscillation at different region of the disk differ slightly by frequency, which causes some smoothening effects, i.e., some oscillating region followed by some region of cancelation.

## 4.3 Interpretation of the Results

To understand the behavior of the shock waves qualitatively, we first note that the shocks discussed here are very strong, i.e., $M_-/M_+ \gg 1$. In this case, it is easily shown that,

$$P_+ = \frac{3}{4}\rho_- v_-^2. \tag{10a}$$

The pre-shock flow has a very low angular momentum and is almost freely falling. Thus, as the shock,

$$v_-(x_s) = \frac{1}{x_s^{1/2}} \tag{10b}$$

and,

$$v_+(x_s) = \frac{1}{4}v_-(x_s) = \frac{1}{4}\frac{1}{x_s^{1/2}}. \tag{10c}$$

The density of the post-shock flow is

$$\rho_+ \sim 4\rho_- = \frac{4\dot{M}}{x_s^{1/2}} \tag{10d}$$

and from Eqn. (10a), the sound speed at the post-shock position is obtained as,

$$a_+(x_s) = \frac{1}{4}(\frac{5}{x_s})^{1/2}. \tag{10e}$$

The post-shock Mach number of the flow is,

$$M_+ = \frac{v_+}{a_+} = 0.2^{1/2} \sim 0.45. \tag{10f}$$

at the shock. The location of the shock is determined by employing the Mach number relation (Eqn. 5) by noting that the shock invariant constant for a strong shock $M_- \to \infty$ is given by

$$C = 1/\gamma^2 = 0.36. \tag{10g}$$

In all our solutions we do see that the post-shock Mach number is $\sim 0.45$ (e.g., Fig. 2) and at the shock location $C \sim 0.36$ as described above.

As the shock is perturbed and propagates outwards, it heats the post-shock flow to a higher temperature since the relative velocity (which is thermalized in the post-shock region) between the shock and the incoming flow is higher (Fig. 3a). As a result, the cooling rate goes up and the outward motion of the shock stops when the flow is sufficiently cool. The post-shock pressure drops and the shock returns towards the black hole. The relative motion between the shock and the post-shock flow decreases and the temperature in the post-shock region drops. The lower pressure in the post-shock region is unable to balance the pre-shock ram pressure and the shock collapse continues till the centrifugal barrier is sufficient to hold the flow. The shock then overshoots this region and bounces from higher pressure region and the cycle continues. In order to have a oscillatory behavior, the post-shock region must be able to cool in a time-scale $t_{cool}$ comparable to the advection time-scale $t_{adv}$. If $t_{cool} << t_{adv}$, the flow quickly adjusts to the cooler state and if $t_{cool} >> t_{adv}$ the flow advects all the thermal energy. Only when these time scales are comparable, the flow tries to constantly adjust itself, and the oscillation is seen. Thus, the time period of oscillation $t_{oscn}$ is roughly comparable to the cooling time $t_{cool} \sim t_{adv}$. In above simulation, we have chosen the accretion rate in a manner that these time scales are comparable when $\alpha = 0.5$. For this particular accretion accretion rate, $t_{cool} << t_{adv}$, when $\alpha = 0.75$ and the flow adjusts immediately to the new solution. As a result there is no oscillation. For $\alpha = 0.4$, for the same accretion rate, no transonic accreting solution is possible. Here, the cooling time is large and the shock has to move outwards by a large amplitude (and for a longer time) so as to satisfy $t_{cool} \sim t_{oscn}$. The amplitude of oscillation is proportional to the thermal energy content of the flow.

To compare the time scales in our problem, we compute

$$t_{adv} = \int_{x_s}^{1} dx/v$$

and the cooling time due to power-law cooling

$$t_{cool} = \int_{x_s}^{1} d\left(\frac{e}{\zeta_{1/2}\rho_+ e_+^{\alpha}}\right)$$

for various $\alpha$. For $\alpha = 0.5$, $t_{adv} = 7.5 \times 10^5$s and $t_{cool} = 1.35 \times 10^6$s. The time period of oscillation (Fig. 1) is found to be $t_{oscn} = 1.5 \times 10^6$s. Thus, $t_{cool} \approx t_{oscn}$ and $t_{adv}$ is also similar within a factor of 2. On the other hand, for $\alpha = 0.75$, we compute, $t_{adv} = 1.03 \times 10^6$s and $t_{cool} = 3 \times 3$s. Because of the disparity in time scales, no oscillation is seen. For $\alpha = 0.4$, cooling is very high and no transonic solution exists with the accretion rate chosen. The infall time-scale ($t_{adv} \sim 4x_s^{3/2}$) from $x_s \sim 42$ to $x_s \sim 15$ is given by $t_{adv} = 865$units or $t_{adv} \sim 8.6 \times 10^5 M/10^8 M_\odot$s roughly the same time in which half the oscillation is completed (Fig. 1). Thus, the oscillation is purely driven by the centrifugal barrier and the cooling effects: Centrifugal barrier determines the minimum shock location, whereas the amplitude of oscillation is determined by the time scales of various cooling processes.

## 5. CONCLUDING REMARKS

In this paper, we have discovered an important new physical effect in the context of the black hole accretion physics. We showed that if the disk parameters are right enough so that the infall timescale of accretion matter from the shock, roughly agrees with the cooling timescale in the post-shock region, then the shock exhibits a radial oscillation around the mean shock location with the time period as that of the cooling time scale. We have provided examples of such oscillation as well as the damping of the oscillation when one is farther away from 'resonance' behavior. We used a power law cooling process, such as bremsstrahlung, to illustrate our claim. For a given accretion rate, there appears to be a cut-off exponent $\alpha_c$ such that the oscillation is present when $\alpha < alpha_c$, however, it is possible that no 'universal' $\alpha_c$ exists. If the shock forms for a given disk parameter, there could be always some $\alpha$ for which the oscillation of shock may be seen.

Similar phenomenon has been reported in accretion processes onto compact stars which have 'hard' boundaries such as white dwarfs (Langer, Chanmugam & Shaviv, 1981). It is found that $\alpha_c \approx 0.6$ (Imamura, Wolff & Durisen, 1983). However, in these systems the location of the shock may form only very close to the star surface (unlike in the case of a black hole accretion where one can adjust the angular momentum to obtain the shock location far away as well), and therefore there may be a fixed $\alpha_c$ as there is little room to vary the infall time scale. Secondly, we believe that the oscillation should take place for any cooling processes (not only with power law cooling process) provided the time scales agree. In other words, if an oscillation is seen for a given cooling process, it may disappear in presence of some other cooling process if the flow are not changed. This may be the reason for the disappearance of oscillation of Chanmugam, Langer & Shaviv (1985) when synchrotron cooling was added.

The time period of oscillation is found to be of the order several hundred light crossing time of the black hole and therefore could range from milliseconds to months depending upon the mass of the black hole. The amplitude of the emitted radiation, mostly in regions extreme ultraviolet to $\gamma$ rays, should be modulated by as much as ten percent for accretion solution. Because the cool and hot matter co-exist side by side close to a shock, correlated variability is expected in these systems, which may have been observed in some cases as mentioned in the Introduction. The accretion rates needed are reasonable and we expect such phenomenon to be seen in many more systems, particularly in active galaxies where low angular momentum accretion may be taking place. The oscillation frequency only indirectly depends on viscosity since the shock location is affected by viscosity (C90, CM95). In several galactic black hole candidates, such as GS339-4 and GS1124-68, QPO behavior is seen (Dotani, 1992). These oscillations are of $3-8$Hz and the amplitude modulation could be as high as $5-10$ percent depending upon observed energy band. We know of no other physical explanation which can satisfactorily explain so large amplitude variation. Within the framework of our model, if the shock is located at $x_s = 100$, $t_{oscn} \sim t_{adv} = 0.2s$ ($\nu_{QPO} = 5$ for a $M = 5M_\odot$ black hole. For a massive black hole, $M = 10^8 M_\odot$, the time scale of QPO behavior could similarly vary from a few hours to a few days. A firm confirmation of this oscillation may be the test of accretion onto a very compact object through a shock wave. Exactly similar mechanism may be responsible for QPOs in disks around neutron stars as well. How long the oscillation will last depends on how sharp the resonance is. In a dwarf nova type outburst, accretion rate is swept through a large range, and therefore it is expected that in some range the resonance will occur. In white dwarfs candidates, such as SS Cygni, VW Hyi, and U Gem oscillations of $7-11$s (Mauche, Raymond & Mattei, 1995), $\sim 14$s (Van der Woerd et al, 1987) and $25-29$s (Córdova et al. 1984) respectively have been seen. Another prediction of the model is that the time-period of oscillation ($\approx$ cooling time) should decrease (roughly linearly) as the accretion rate is increased. Similarly, oscillation may disappear completely if the accretion rate is out of the resonance width. These are clearly reported in recent observations (Mauche, 1995). Thus, we believe that the observational results may have agreed with our resonance model of oscillation.

If the hard X-rays in galactic black hole candidates are actually produced in the post-shock region (Chakrabarti & Titarchuk, 1995), and the QPO oscillation is due to the oscillation of these shock waves, we see virtually no difference in the *qualitative* properties between the spectra of black hole candidates and that of neutron stars. The quantitative difference will come only from the presence (black hole) or absence (neutron star) of the inner sonic point and thus from flow very close to the horizon. In other words, even if the flow becomes sub-Keplerian close to the black hole with very little radiation efficiency (Abramowicz et al. 1988, C89, C90, Chen & Taam, 1993, Chakrabarti & Titarchuk, 1995) it still finds a way, namely, through the shocks (C89, C90, CM95), to dissipate most of its energy at this 'boundary layer'.

We have employed Smoothed Particle Hydrodynamics to illustrate this important time-dependent behavior of the flow. We observe that the numerical results agree with the analytical solution very accurately thus removing some of the misgivings attributed to SPH simulations. We have captured, quite successfully, very strong shock waves where Mach number jumps from $M_- \sim 100$ to $M_+ \sim 0.45$. We therefore believe that SPH could be a valuable tool to study astrophysical processes provided adequate care is taken to guarantee accuracy in interpolation method, which is accomplished here by the choice of very small particle size.

We have studied the behavior of the accretion flow by considering only very simple physical processes such as compressional heating and bremsstrahlung cooling. The phenomenon that we discover, however, need not be confined to these systems alone. The oscillating behavior is due to generic physical processes such as heating and cooling and should be present even in complex situations where more than one heating and cooling mechanisms, such as Comptonization, line cooling etc. are present. We have shown in our simulation of radiative shocks in thick disks, that the oscillations are smoothed out to some degree and perhaps more closely behave like QPOs than periodic oscillations. In these simulations we show the existence of the hot corona which might be responsible for the hard component observed in galactic and extragalactic black hole candidate spectra.

After completion of this work, we became aware of a very recent work of a purely hydrodynamic simulation (Ryu et al, 1995) of a non-accreting, quasi-spherical flows around a Newtonian star. This simulation also shows some type of oscillation even when no cooling is added. These results seem to be due to the presence of a strong centrifugal barrier of a non-accreting high angular momentum flow. We have been able to reproduce these oscillations when both the cooling and the accretion are absent. In presence of accretion these oscillations go away and a steady state shock is achieved as in Molteni, Lanzafame & Chakrabarti (1994). Our present paper shows that to obtain a sustained periodic oscillation in an accreting disk, one requires to include cooling processes as in a realistic system.

SKC would like to thank the hospitality of the Landessternwarte, and the Institute of Theoretical Astrophysics, University of Heidelberg where a part of the work was carried out. DM would like to thank the hospitality of the Institute of Theoretical Astrophysics, University of Heidelberg. Work of SKC at the Goddard Space Flight Center, is supported by a Senior Research Associateship award of the National Academy of Science.

Figure Captions

Fig.1: Comparison of the time dependent shock locations as functions of time for $\alpha = 0.4$ (long-dashed line), $\alpha = 0.5$ (solid line) and $\alpha = 0.75$ (short-dashed line). Same accretion rate and angular momentum is used at the outer edge for all the three cases.

Fig. 2: Comparison of the time dependent (long-dashed, short-dashed and dotted lines) and the steady state solutions (solid line) at various phases of oscillation when bremsstrahlung emission is considered. The time dependent solution oscillates close to the analytic solution. The dash-dotted curve represents the solution without any cooling effects.

Fig. 3(a-b): The variations of (a) temperature (in units of $10^9$K) and (b) total bremsstrahlung loss (in arbitrary units) in various phases of oscillation.

Figure 4: The loss of energy due to cooling as a function of time. Solid curve represents sinusoidal bremsstrahlung loss ($\alpha = 0.5$) and the dashed curve represents the damped oscillation for $\alpha = 0.75$ cooling.

Fig. 5(a-d): Results of numerical simulations in three dimensional axisymmetric flows. In the upper half of each Figure, we have contours of constant Mach number $M = 1$ (thick curves) and the velocity vectors and in the lower half we show the density contours (in units of density at the outer edge). (a) No cooling and (b-d) Post-shock corona at different phases of oscillations when the bremsstrahlung is included.

Fig. 6: Oscillation of the net loss of energy due to bremsstrahlung as a function of time from a thick corona.